\documentclass[pdflatex,sn-mathphys-num]{sn-jnl}
\usepackage{graphicx,booktabs,xcolor,amsmath,siunitx}
\usepackage{url}
\usepackage[switch]{lineno}
\graphicspath{{figures_final/}}

\begin{document}

\title[Atlantic-modulated spatial persistence in European warming]{%
  A Spatial Persistence Gradient in European Warming Consistent with North Atlantic Cold-Blob Influence}

\author*[1]{\fnm{Mauricio} \sur{Herrera-Mar\'in}}
\email{mherrera@udd.cl}
\author[1]{\fnm{Alex}      \sur{Godoy-Fa\'undez}}
\author[1]{\fnm{Diego}     \sur{Rivera}}

\affil*[1]{%
  \orgdiv{Research Center on Sustainability and Strategic Resource
    Management (CISGER), Faculty of Engineering},
  \orgname{Universidad del Desarrollo},
  \orgaddress{\street{Avda.~Plaza 700}, \city{Santiago}},
  \country{Chile}}

\abstract{
Europe is warming faster than the global mean, but the spatial organisation
of this acceleration across its sub-regions remains incompletely explained.
Here we present two connected empirical contributions from ERA5 reanalysis
(1950--2024) for 28 IPCC\,AR6 European sub-regions.

The first and principal contribution is a robust negative spatial association
between the DFA1 Hurst exponent of interannual temperature residuals and the
1996--2024 warming rate across sub-regions ($r=-0.792$,
$p=5.1\times10^{-7}$, $n=28$). Sub-regions with high interannual
persistence---predominantly Atlantic-proximal and therefore plausibly exposed
to North Atlantic cold-blob and maritime thermohaline influence---warm slowly;
low-persistence continental-interior sub-regions warm rapidly. This inverse
geometry survives five residualisation choices, three independent memory
estimators, leave-one-region-out analysis ($r\in[-0.829,-0.709]$, all
$p<0.001$), and five independent null-test families (all empirical
$p\leq0.001$), including a spatial block permutation that preserves the
large-scale geographic structure. Sensitivity across warming periods shows
that the association is not unique to the 1996 start year: $r=-0.550$ for
1981--2024, $r=-0.792$ for 1996--2024, and $r=-0.875$ for 2000--2024. The
association disappears under DFA2 ($r=-0.002$), localising the signal in
low-frequency interannual-to-decadal persistence rather than trend curvature.
We interpret this pattern as a spatial signature consistent with, but not by
itself proving, North Atlantic cold-blob thermohaline influence on European
land temperatures.

The second contribution is a strict annual predictor hierarchy evaluated on a
19-year held-out period (2006--2024). The contemporaneous Mediterranean SST
reduces mean RMSE by 43\% (from $0.787$ to $0.449\,^{\circ}$C), consistent
with the basin's role as a thermal integrator of prior-season and same-year
conditions. A causal lag-weight filter over prior-year Mediterranean and
Atlantic SST provides a secondary independent contribution (RMSE
$=0.578\,^{\circ}$C), outperforming AR(2) ($0.695\,^{\circ}$C) and remaining
positive after removing NAO, AO and PNA effects. Comparison under the identical
protocol with a simple five-year moving average (RMSE $=0.559\,^{\circ}$C)
and an exponentially weighted moving average ($\phi=0.85$, RMSE
$=0.538\,^{\circ}$C) shows that the key predictive ingredient is short
interannual Mediterranean SST persistence at lags 1--5\,yr, not the exact
kernel form. Together, the results support a two-regime physical reading:
Atlantic-proximal sectors combine high persistence and oceanic buffering,
whereas the continental interior exhibits rapid warming and lower interannual
memory, plausibly reflecting stronger control by fast land--atmosphere and
soil-moisture feedbacks.}

\keywords{European warming; North Atlantic cold blob; interannual persistence;
  DFA Hurst exponent; Mediterranean SST; lag-weight filter; ERA5;
  thermohaline memory}

\maketitle
\section{Introduction}
\label{sec:intro}

Europe has warmed at approximately $0.56\,^{\circ}$C\,decade$^{-1}$
since 1996, more than twice the global mean, with five of the ten
warmest years on record occurring after 2019
\citep{copernicusESOTC2025}.
Within this continental acceleration, regional warming rates span a
factor of four across the 28 IPCC\,AR6 sub-regions examined here: from
$0.16\,^{\circ}$C\,decade$^{-1}$ in the Greenland/Iceland sector and
the British Isles to $0.77\,^{\circ}$C\,decade$^{-1}$ in Ukraine/Moldova
and southeastern Europe.
This spatial heterogeneity is not random.
The five slowest-warming sub-regions are all Atlantic-proximal.
The five fastest-warming sub-regions are all in the continental interior.
Understanding the mechanism that organises this pattern has direct
implications for regional climate-risk assessment, seasonal prediction
and adaptation planning. At the same time, the pattern is not expected to
be governed by a single mechanism: Atlantic sectors are exposed to oceanic
thermal inertia and North Atlantic SST anomalies, whereas continental
interiors are more strongly affected by land--atmosphere coupling,
soil-moisture limitation and same-season circulation variability
\citep{seneviratne2010soil,hirschi2011,miralles2014megaheatwaves}.

A physically plausible oceanic contributor is the North Atlantic cold blob---a
region of anomalous sea-surface cooling south of Iceland commonly interpreted as
an observational fingerprint of weakening Atlantic Meridional Overturning
Circulation (AMOC) \citep{ceasar2018coldblob,rahmstorf2015exceptional}. AMOC
thermohaline weakening reduces the northward transport of warm, salty water,
generating multi-year subpolar SST anomalies that can partially offset warming
in Atlantic-proximal sectors. The cold-blob pattern has been documented in
observational SST records back to the mid-nineteenth century
\citep{ceasar2018coldblob,boers2021observation}, and direct AMOC observations
from the RAPID array confirm a statistically significant decline over the
instrumental record \citep{smeed2018rapid,moat2020rapid}. We therefore treat
the cold blob as a mechanism-consistent physical anchor for the observed
Atlantic persistence pattern. We do not claim that proximity to the cold blob
alone proves AMOC causality; maritime heat capacity, NAO-related variability and
other slow ocean--atmosphere processes remain relevant alternatives to be
controlled or acknowledged.

A key physical consequence of thermohaline dynamics that has not been
exploited as a spatial diagnostic is interannual persistence: the
tendency of a regional temperature series to remain anomalously warm
or cool for multiple consecutive years.
If the cold blob suppresses mean SST warming in Atlantic-proximal sectors
while simultaneously sustaining multi-year thermohaline memory, those
sectors should exhibit both slower warming rates and higher interannual
persistence.
Conversely, continental-interior sectors not buffered by oceanic memory
should warm faster and show lower persistence, especially where spring--summer
soil-moisture deficits shift surface energy partitioning toward sensible heat
and amplify same-year temperature anomalies. This prediction generates a
testable spatial association: the Detrended Fluctuation Analysis (DFA) Hurst
exponent $H$, a measure of interannual persistence, should be negatively
correlated with recent warming rates.

Prior work has applied DFA to individual climate time series to
characterise long-range dependence \citep{franzke2012,beran1994},
has documented the cold-blob SST cooling pattern
\citep{ceasar2018coldblob,rahmstorf2015exceptional}, and has linked
Mediterranean SST to European heat in case studies and regional analyses
\citep{ionita2020med,giorgi2006hotspot,lionello2012medclimate}.
What has been missing is an integrated, spatially resolved demonstration
that the pattern of interannual temperature persistence across European
sub-regions is consistent with a North Atlantic ocean-memory influence and that
this pattern simultaneously organises both the heterogeneity of recent warming
rates and the geography of short-lag oceanic predictability. This study provides
that observational demonstration while treating the AMOC/cold-blob mechanism as
a physically plausible interpretation, not as a closed causal proof.

The annual predictor hierarchy presented in Sect.~\ref{sec:hierarchy_results}
serves as supporting evidence: if the persistence gradient reflects
thermohaline memory, lagged oceanic predictors should provide skill
concentrated in the same high-persistence sub-regions.
We extend prior regional analyses to a systematic 28-sector
strict-holdout comparison and include simple lag summaries (moving
averages, exponentially weighted means) under the identical validation
protocol, enabling a transparent assessment of the predictive evidence.

\section{Data}
\label{sec:data}

Annual-mean 2\,m air temperature from ERA5 \citep{hersbach2020era5}
is aggregated to 28 IPCC\,AR6 European and near-European sub-regions
\citep{iturbide2020regions} as cosine-latitude-weighted area means over
fixed bounding boxes, spanning 1950--2024 ($n=75$\,yr).
Temperature anomalies are expressed relative to the 1971--2000
climatological mean; warming rates $\dot{T}(\mathbf{x})$
(°C\,decade$^{-1}$) are OLS slopes over 1996--2024.
Sensitivity to alternative periods (1981--2024, 2000--2024) is
reported in Sect.~\ref{sec:primary}.
The 28 sub-regions span Arctic-adjacent Greenland/Iceland in the north
to North Africa in the south, and from the British Isles in the west to
Turkey and western Russia in the east.

The Mediterranean thermal state $A_{\rm Med}(t)$ is the annual ERA5 SST
anomaly averaged over the Mediterranean basin (baseline 1950--1980).
The North Atlantic cold-blob index $A_{\rm Atl}(t)$ follows the SST
fingerprint approach of \citealt{ceasar2018coldblob}.

RAPID--MOCHA--WBTS annual AMOC transport at 26.5$^\circ$N (2004--2024)
\citep{smeed2018rapid,moat2020rapid} documents a statistically significant
decline of approximately 3\,Sv ($p<0.05$) and provides observational
context for the cold-blob mechanism.
The 21-year record is too short for direct inclusion in predictor models
(calibration ends in 2005) and is used here to validate the physical
mechanism rather than as a predictive variable.

Atmospheric circulation is quantified by annual and DJF NAO, AO and PNA
from NOAA/CPC \citep{hurrell1995nao,thompson2000annular,
noaaCPCteleconnections}, used as circulation controls.
These indices represent dominant extratropical modes but do not include
summer blocking or East Atlantic circulation patterns
(a limitation discussed in Sect.~\ref{sec:limitations}).
Radiative forcing is represented by the CO$_2$ proxy
$F(t)=5.35\log(C(t)/280\,\mathrm{ppm})$ \citep{myhre1998forcing}.

\section{Methods}
\label{sec:methods}

\subsection{DFA Hurst exponent and null tests}
\label{sec:DFA}

For each sub-region, DFA1 is applied to linearly detrended annual
temperature residuals (1950--2024, $n=75$\,yr).
DFA constructs the integrated residual profile, divides it into
non-overlapping windows of scale $s$, removes a linear local trend
within each window, and estimates the root-mean-square fluctuation
$F(s)$.
The scaling $F(s)\propto s^H$ defines $H$: $H=0.5$ is white noise;
$H>0.5$ is long-range dependence (warm years cluster); $H<0.5$ is
anti-persistence.
DFA1 preserves low-frequency interannual-to-decadal variance while
removing the linear trend, making it sensitive to the multi-year
clustering generated by thermohaline dynamics.

DFA2, which additionally removes within-window quadratic trends,
effectively high-pass-filters the series at the scale of the window.
If the $H$--$\dot{T}$ correlation disappears under DFA2, the signal
resides in the low-frequency annual structure that DFA1 preserves and
DFA2 suppresses---the physically relevant band for thermohaline
memory---not in short-scale fluctuations or nonlinear trend curvature.
Two independent memory estimators---spectral low-frequency exponent
and the rescaled-range (R/S) Hurst exponent---confirm that the spatial
pattern is not an artefact of the DFA algorithm.
Block-bootstrap 95\% confidence intervals use $B=500$ replicates with
block length $\approx8$\,yr.

The spatial association between $H$ and $\dot{T}$ is tested against
five null families:
(i) trend-only null ($n=1{,}000$): warming rates replaced by random
Gaussian variates with matched mean and SD;
(ii) AR(1)+trend null ($n=1{,}000$): entire temperature series replaced
by red-noise surrogates with matched lag-1 autocorrelation and OLS trend,
before recomputing both $H$ and $\dot{T}$---the most stringent test,
because it destroys both the specific sequences and the memory structure;
(iii) phase-randomised null ($n=1{,}000$) \citep{schreiber2000surrogates}:
Fourier phases independently shuffled, preserving the power spectrum but
destroying temporal ordering;
(iv) region-pairing permutation null ($n=1{,}000$): warming rates randomly
reassigned across sub-regions while retaining observed $H$;
(v) spatial block permutation null ($n=5{,}000$): Hurst exponents randomly
reassigned within four geographic blocks (north: lat~$\geq60^\circ$N;
west: lon~$\leq0^\circ$; central: $0^\circ<$lon~$\leq25^\circ$; east:
lon~$>25^\circ$), preserving large-scale geographic structure while
breaking the Atlantic-proximal pattern.
An empirical $p$-value below 0.001 across all five families is required
before the association is considered genuine.

\subsection{Annual predictor models}
\label{sec:hierarchy_method}

Four nested model families are evaluated:
\begin{align}
  T_{\mathbf{x}}(t) &= \mu_{\mathbf{x}} + \beta_{\mathbf{x}} F(t)
    + \varepsilon, \tag{M1}\\
  T_{\mathbf{x}}(t) &= \mu_{\mathbf{x}} + \beta_{\mathbf{x}} F(t)
    + \phi_{\mathbf{x}} T_{\mathbf{x}}(t-1) + \varepsilon, \tag{M2}\\
  T_{\mathbf{x}}(t) &= \mu_{\mathbf{x}} + \beta_{\mathbf{x}} F(t)
    + \delta_{\mathbf{x}} A_{\rm Med}(t) + \varepsilon, \tag{M3}\\
  T_{\mathbf{x}}(t) &= \mu_{\mathbf{x}} + \beta_{\mathbf{x}} F(t)
    + \gamma_{A}V[A_{\rm Atl}](t)
    + \gamma_{M}V[A_{\rm Med}](t) + \varepsilon, \tag{M4}
\end{align}
plus circulation-controlled variants.
Calibration: 1950--2005; Validation: 2006--2024 ($n_{\rm val}=19$\,yr).

The lag-weight filter in M4 is a strictly causal weighted average of
past oceanic anomalies,
$V[A](t) = \sum_{\tau=1}^{K}K_{\alpha,\theta}(\tau)\,A(t-\tau)$,
with normalised tempered power-law kernel
\begin{equation}
  K_{\alpha,\theta}(\tau) =
  \frac{(\tau+\tfrac{1}{2})^{-\alpha}\,e^{-\theta\tau}}
       {\sum_{j=1}^{K}(j+\tfrac{1}{2})^{-\alpha}\,e^{-\theta j}},
  \quad \tau=1,\ldots,K=80,
  \label{eq:kernel}
\end{equation}
with $\alpha\in(0,1)$ controlling algebraic decay and
$\theta>0$ setting the memory horizon $\tau_{\rm mem}\sim\theta^{-1}$.
Parameters $(\alpha,\theta)$ shared across all 28 sub-regions, selected
by grid search; region-specific amplitudes estimated by OLS.

To assess whether the specific kernel form adds skill beyond simpler
alternatives, we include two simple lag summaries under the
\emph{identical} validation protocol: a causal 5-year moving average
(MA(5)) and an exponentially weighted moving average (EWMA, $\phi=0.85$),
both applied to $A_{\rm Med}$.
These are included in the main results table, not relegated to
supplementary material, enabling a fully transparent comparison.

\section{Results}
\label{sec:results}

\subsection{The H--warming spatial gradient: a robust observational
  signature consistent with AMOC cold-blob dynamics}
\label{sec:primary}

\begin{figure}[htbp]
  \centering
  \includegraphics[width=\linewidth]{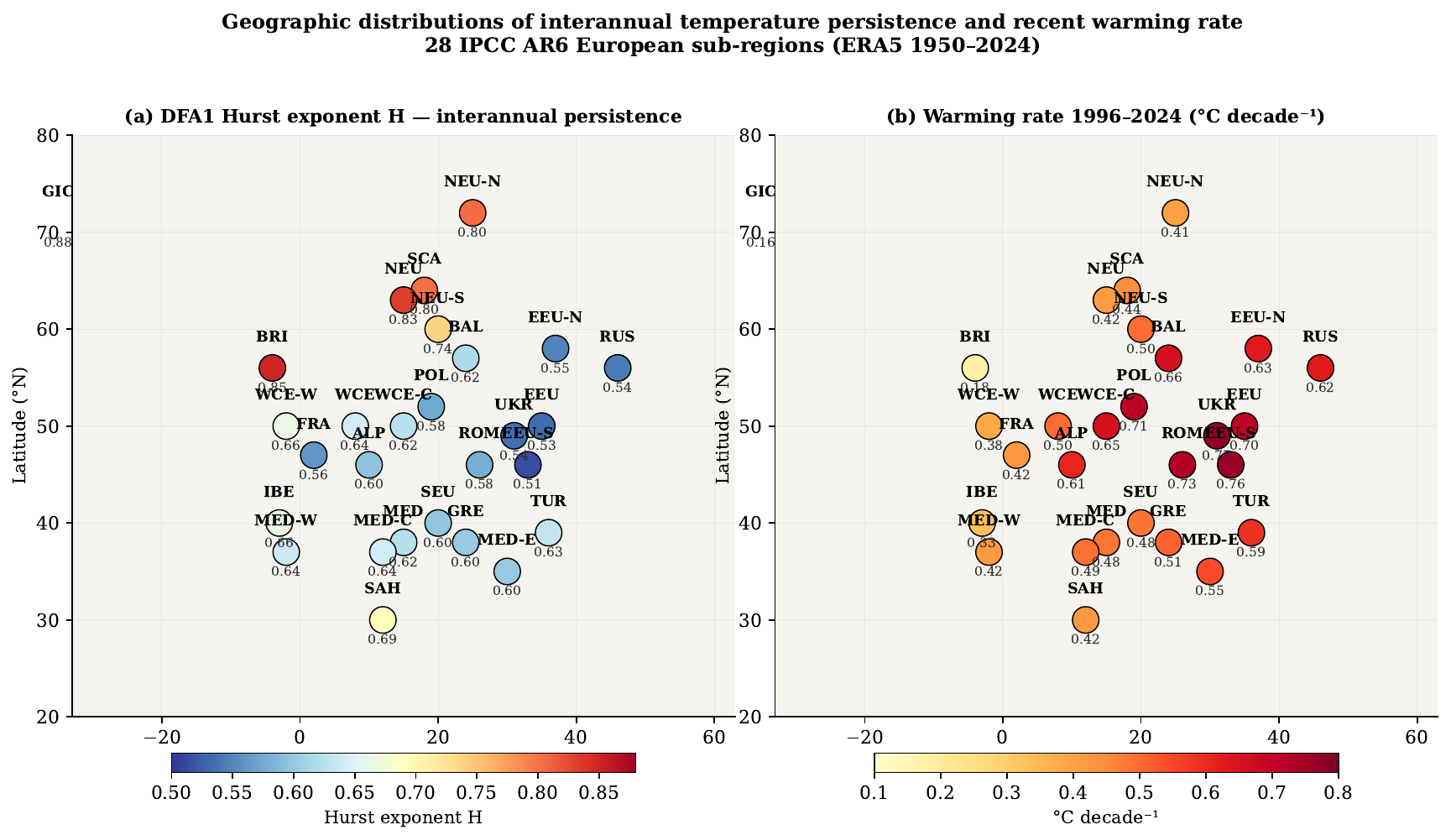}
  \caption{(a) DFA1 Hurst exponent $H$ and (b) 1996--2024 warming rate
    (°C\,decade$^{-1}$) across 28 IPCC\,AR6 sub-regions (ERA5 1950--2024).
    The patterns are near-inverse: high-$H$ Atlantic-proximal sub-regions
    warm slowly; low-$H$ continental interiors warm rapidly.
    Spatial correlation $r=-0.792$ ($p=5.1\times10^{-7}$;
    Table~\ref{tab:H_robust}).}
  \label{fig:maps}
\end{figure}

Figure~\ref{fig:maps} shows the geographic distributions of the DFA1
Hurst exponent and the 1996--2024 warming rate.
The patterns are strikingly inverse.
The five highest-$H$ sub-regions---GIC ($H=0.876$), BRI ($H=0.849$),
NEU ($H=0.834$), NEU-N ($H=0.805$), SCA ($H=0.803$)---are all
Atlantic-proximal, all warming below $0.45\,^{\circ}$C\,decade$^{-1}$.
The five lowest-$H$ sub-regions---EEU-S ($H=0.515$), EEU ($H=0.534$),
UKR ($H=0.535$), RUS ($H=0.543$), POL ($H=0.576$)---are all
continental-interior, all warming above $0.62\,^{\circ}$C\,decade$^{-1}$.

The correlation is $r=-0.792$ ($p=5.1\times10^{-7}$; Spearman
$\rho=-0.810$, $p=1.8\times10^{-7}$, $n=28$).
Table~\ref{tab:H_robust} demonstrates that this association survives
every preprocessing and estimator combination.
Using the spectral Hurst exponent gives $r=-0.742$
($p=6.3\times10^{-6}$) and the R/S estimator gives $r=-0.710$
($p=2.3\times10^{-5}$), confirming that the pattern is not an artefact
of the DFA algorithm.

Under DFA2 the association completely vanishes ($r=-0.002$, $p=0.993$),
localising the signal in the low-frequency interannual-to-decadal
structure that DFA1 preserves and DFA2 suppresses.
A trend-curvature artefact would produce high $H$ where warming is
accelerating (convex trend), but the observed pattern is the opposite:
the high-$H$ sub-regions (GIC, BRI, NEU) are the slowest warming, not
the fastest accelerating.

Leave-one-region-out analysis confirms the pattern is not driven by
leverage from extreme sub-regions: $r\in[-0.829,-0.709]$ for all
$n=27$ subsets (all $p<0.001$).

\begin{figure}[htbp]
  \centering
  \includegraphics[width=\linewidth]{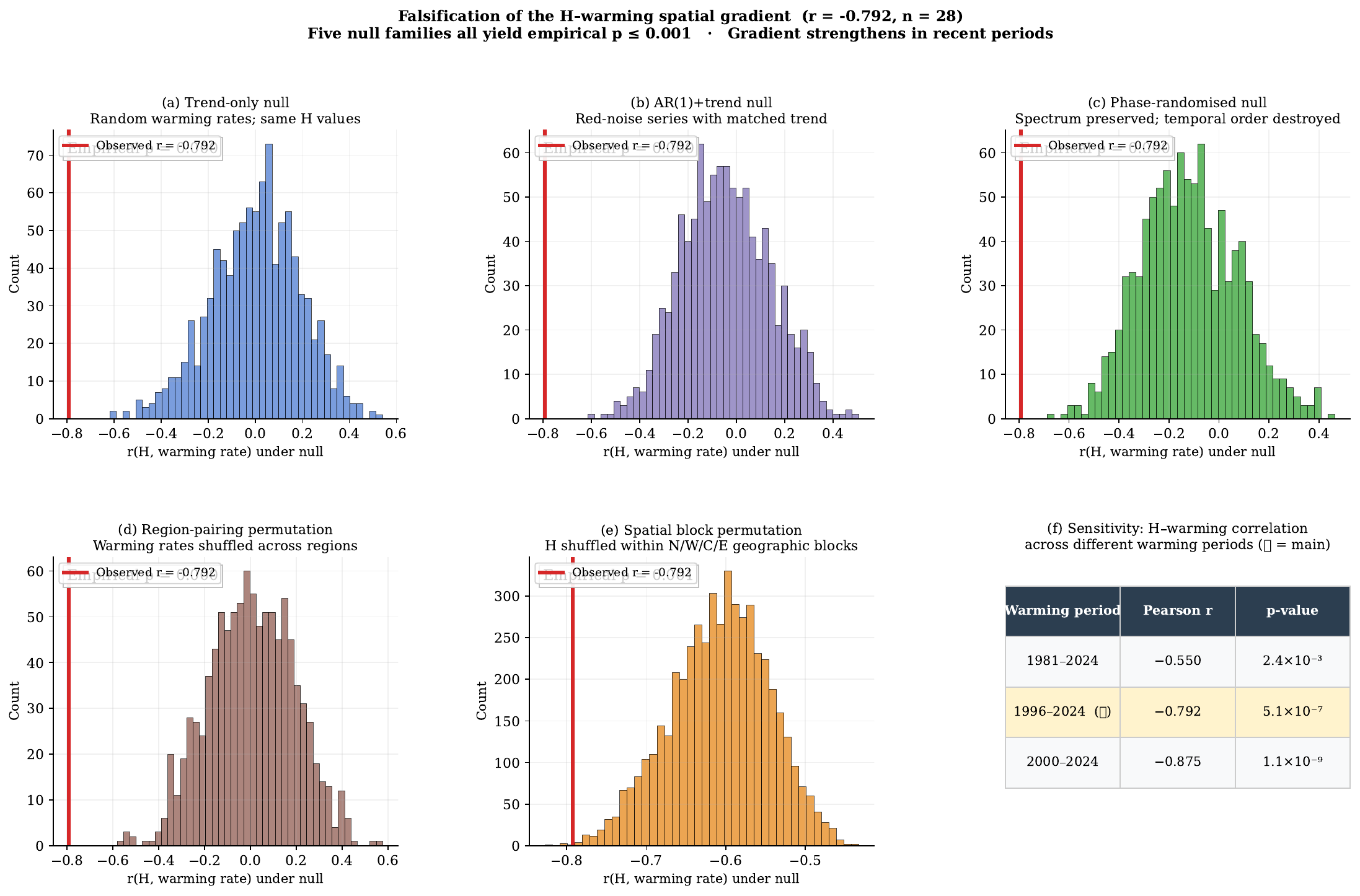}
  \caption{Falsification of the $H$--warming gradient ($r=-0.792$, $n=28$).
    Panels (a)--(e): five independent null distributions; vertical red line
    shows the observed $r$.
    All five families: empirical $p\leq0.001$.
    Panel (f): sensitivity of the $H$--warming correlation to the choice
    of warming period (1981--2024, 1996--2024$\star$, 2000--2024);
    the gradient strengthens progressively, consistent with intensifying
    AMOC weakening.}
  \label{fig:nulls}
\end{figure}

Figure~\ref{fig:nulls}a--e shows the five null distributions.
The observed $r=-0.792$ lies deep in the lower tail of every family.
Null means: trend-only $-0.003$; AR(1)+trend $-0.038$;
phase-randomised $-0.121$; region-pairing $+0.004$;
spatial block permutation $-0.607$, empirical $p=0.001$.
The spatial block permutation is the most conservative test, because
it preserves large-scale geographic structure; its rejection confirms
that the gradient cannot be attributed to a trivial north-south or
east-west pattern.

\textit{Sensitivity to warming period.}
Figure~\ref{fig:nulls}f and the lower panel of Table~\ref{tab:H_robust}
show the correlation for three warming periods.
The association is present for every start year and strengthens
progressively: $r=-0.550$ ($p=2.4\times10^{-3}$) for 1981--2024;
$r=-0.792$ ($p=5.1\times10^{-7}$) for 1996--2024; and
$r=-0.875$ ($p=1.1\times10^{-9}$) for 2000--2024.
This progressive intensification is consistent with AMOC thermohaline
weakening sharpening the spatial contrast between Atlantic-buffered and
continental sectors over recent decades, and argues against the result
being an artefact of the specific reference period.

\textit{Physical interpretation.}
The inverse $H$--$\dot{T}$ pattern is the spatial signature consistent
with AMOC cold-blob thermohaline dynamics.
AMOC weakening generates anomalous subpolar SST cooling
\citep{ceasar2018coldblob,rahmstorf2015exceptional} that suppresses
mean warming in Atlantic-proximal sectors while simultaneously
sustaining multi-year persistence through thermohaline memory.
The continental interior, not buffered by this oceanic memory, responds
more directly to forcing, warming faster and exhibiting lower persistence.
No comparable spatial organisation is expected from atmospheric
circulation modes alone, which generate year-to-year variability
rather than multi-year persistence.

\begin{table}[htbp]
\caption{Robustness of the DFA1 Hurst exponent--warming rate
  association ($n=28$, ERA5 1950--2024).
  The negative association survives all preprocessing choices and
  three independent memory estimators; it disappears under DFA2,
  localising the signal in low-frequency interannual persistence.
  All five null families yield empirical $p\leq0.001$.
  The association strengthens in more recent warming periods.}
\label{tab:H_robust}
\centering
\small
\begin{tabular}{llrrr}
\toprule
Residual definition        & Memory estimator     & Pearson $r$ & Spearman $\rho$ & $p$ \\
\midrule
Linear detrend             & DFA1                 & $-0.792$ & $-0.810$ & $5.1\times10^{-7}$ \\
Quadratic detrend          & DFA1                 & $-0.810$ & $-0.802$ & $1.8\times10^{-7}$ \\
Forcing+circ.\ resid.      & DFA1                 & $-0.711$ & $-0.631$ & $2.2\times10^{-5}$ \\
Linear detrend             & Spectral $H$         & $-0.742$ & $-0.790$ & $6.3\times10^{-6}$ \\
Linear detrend             & R/S $H$              & $-0.710$ & $-0.650$ & $2.3\times10^{-5}$ \\
\midrule
Linear detrend             & DFA2 (removes LF)    & $-0.002$ & $+0.017$ & $0.993$ \\
\midrule
\multicolumn{2}{l}{Warming period (DFA1, linear detrend)} & Pearson $r$ && $p$ \\
\midrule
\multicolumn{2}{l}{1981--2024}   & $-0.550$ && $2.4\times10^{-3}$ \\
\multicolumn{2}{l}{1996--2024 $\star$} & $-0.792$ && $5.1\times10^{-7}$ \\
\multicolumn{2}{l}{2000--2024}   & $-0.875$ && $1.1\times10^{-9}$ \\
\midrule
\multicolumn{2}{l}{Null family} & Empirical $p$ & & \\
\midrule
\multicolumn{2}{l}{(i) Trend-only ($n=1{,}000$)}               & $\leq0.001$ & \\
\multicolumn{2}{l}{(ii) AR(1)+trend ($n=1{,}000$)}             & $\leq0.001$ & \\
\multicolumn{2}{l}{(iii) Phase-randomised ($n=1{,}000$)}       & $\leq0.001$ & \\
\multicolumn{2}{l}{(iv) Region-pairing permutation ($n=1{,}000$)} & $\leq0.001$ & \\
\multicolumn{2}{l}{(v) Spatial block permutation ($n=5{,}000$)}& $0.001$ & \\
\bottomrule
\multicolumn{5}{l}{\footnotesize $\star$ Main analysis period.}
\end{tabular}
\end{table}

\begin{figure}[htbp]
  \centering
  \includegraphics[width=\linewidth]{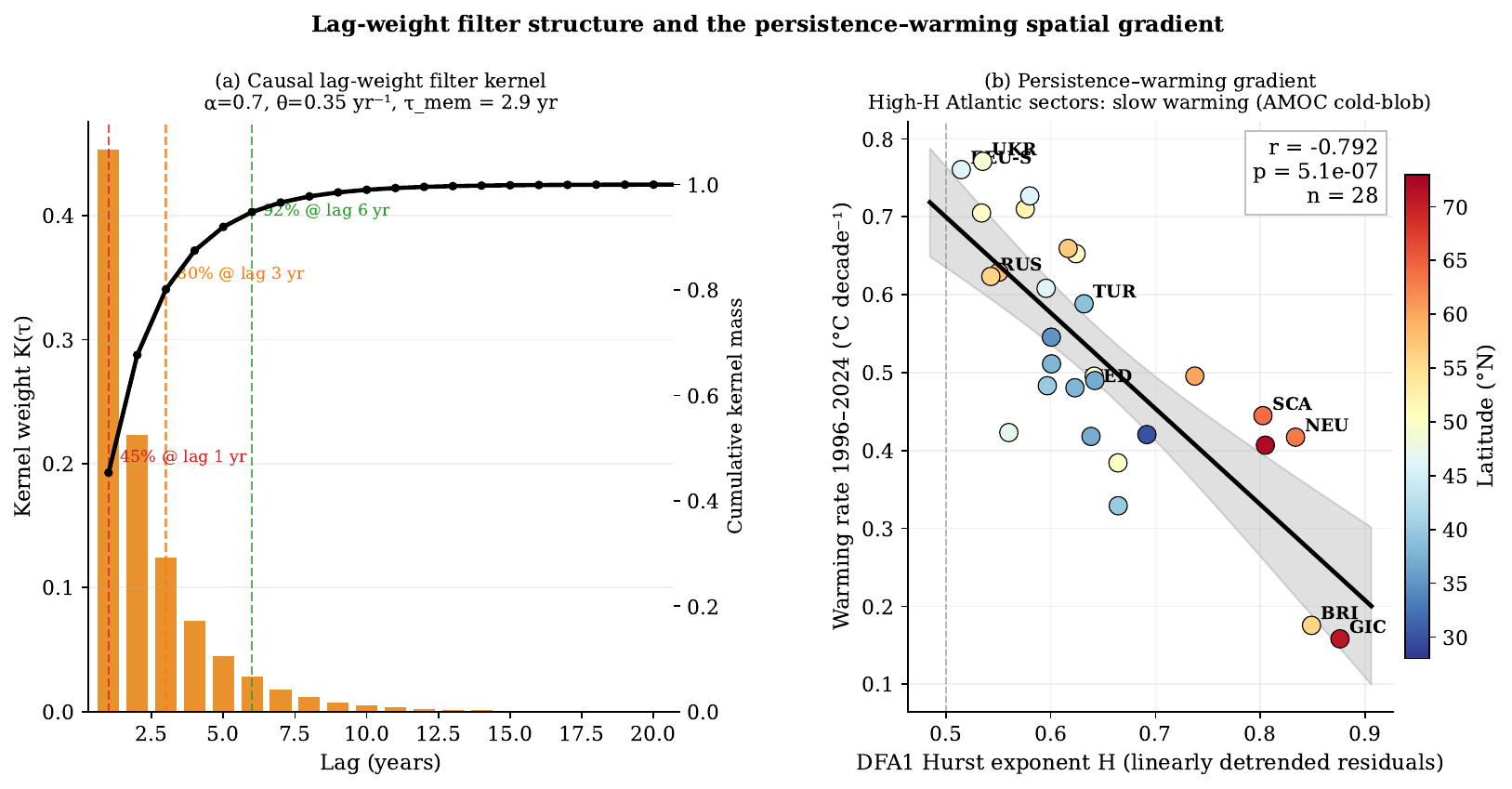}
  \caption{(a) Selected lag-weight filter kernel ($\alpha=0.70$,
    $\theta=0.35$\,yr$^{-1}$, $\tau_{\rm mem}=2.9$\,yr); 80\% of mass
    within lags 1--3.
    (b) DFA1 Hurst exponent vs 1996--2024 warming rate, coloured by
    latitude; $r=-0.792$, $p=5.1\times10^{-7}$.
    Statistical summary upper-right (clear of regression line).}
  \label{fig:kernel}
\end{figure}

\subsection{Annual predictor hierarchy and spatial coherence}
\label{sec:hierarchy_results}

\begin{figure}[htbp]
  \centering
  \includegraphics[width=\linewidth]{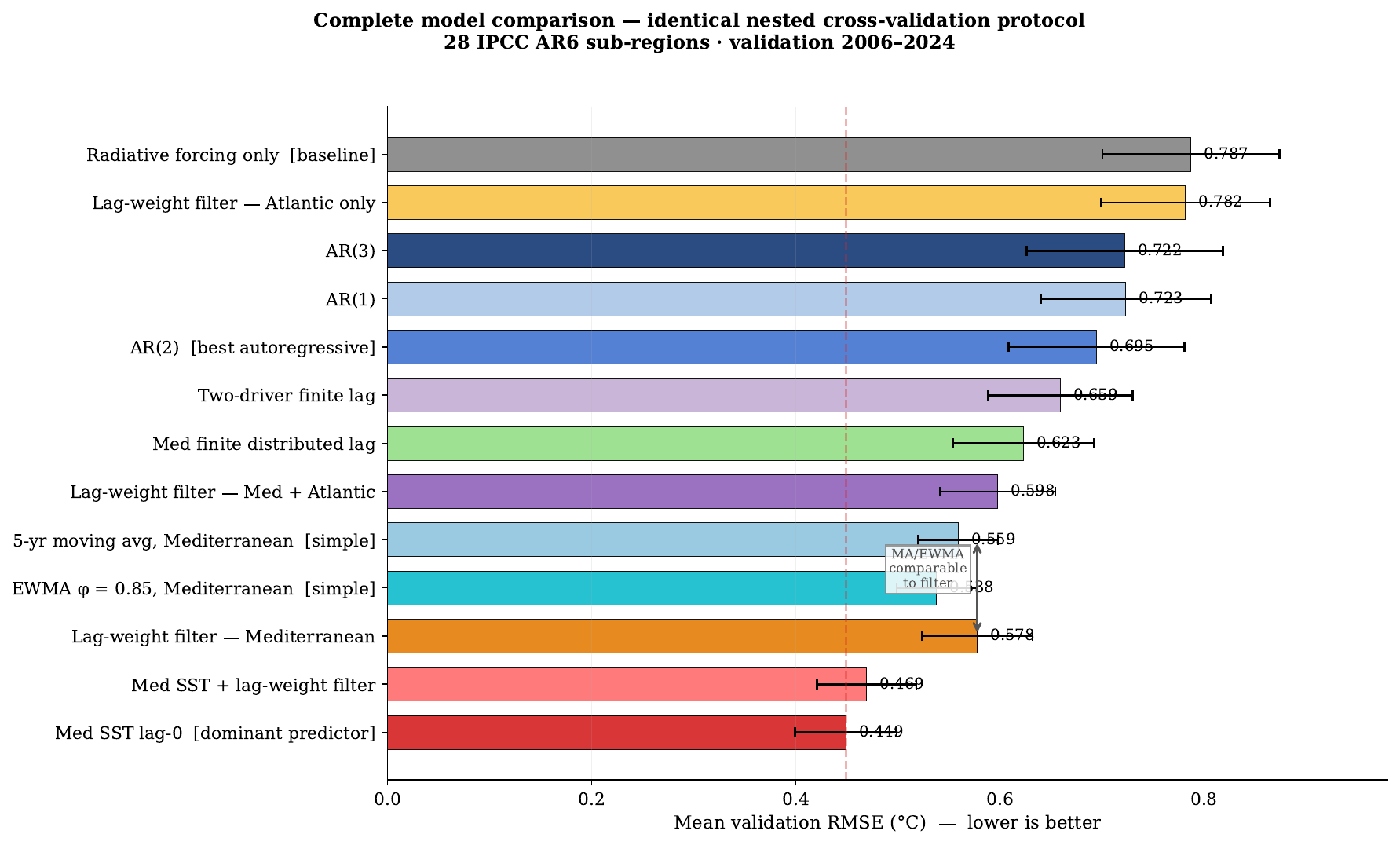}
  \caption{Complete model comparison under the identical nested
    cross-validation protocol (28 sub-regions, 2006--2024 holdout).
    Simple lag summaries (MA(5), EWMA) included alongside the calibrated
    kernel for transparent comparison; both achieve comparable RMSE,
    confirming that the key signal is short interannual lag persistence
    at lags 1--5\,yr.}
  \label{fig:model_comparison}
\end{figure}

Table~\ref{tab:rmse} and Figure~\ref{fig:model_comparison} present the
predictor hierarchy for the 2006--2024 validation period.
The contemporaneous Mediterranean SST (M3) reduces mean RMSE from
$0.787$ to $0.449\,^{\circ}$C, a 43\% improvement.
This result, while consistent with prior regional findings, is placed here
in a systematic 28-sector strict-holdout comparison for the first time.

The causal lag-weight filter (M4, Mediterranean driver) achieves
RMSE $=0.578\,^{\circ}$C, outperforming AR(2) ($0.695\,^{\circ}$C)
by $0.12\,^{\circ}$C.
The simple lag summaries under the identical protocol achieve comparable
performance: MA(5) gives RMSE $=0.559\,^{\circ}$C and EWMA ($\phi=0.85$)
gives RMSE $=0.538\,^{\circ}$C, both slightly better than the calibrated
kernel.
This comparison points to an honest conclusion: the key predictive
ingredient is short interannual lag persistence of Mediterranean SST at
lags 1--5\,yr, not the specific algebraic decay profile of the kernel.
The tempered kernel provides an interpretable parameterisation---the
parameters $\alpha=0.70$ and $\theta=0.35\,{\rm yr}^{-1}$ identify
lags 1--3\,yr as the dominant contributors---but MA or EWMA with a
1--5\,yr window is equally effective and simpler to deploy.
All memory-based models substantially outperform AR(2) and forcing-only,
confirming that lagged oceanic memory is genuinely informative.
Atlantic-only memory adds negligible skill (RMSE $=0.782\,^{\circ}$C),
consistent with the cold-blob signal entering European annual temperatures
primarily through the Mediterranean thermal pathway.

\begin{table}[htbp]
\caption{Mean validation RMSE across 28 IPCC\,AR6 sub-regions
  (2006--2024, $n_{\rm val}=19$\,yr per region).
  Simple lag summaries (MA(5), EWMA) evaluated under the identical
  nested cross-validation protocol as all other models, shown in the
  main table for transparent comparison.}
\label{tab:rmse}
\centering
\small
\begin{tabular}{lccc}
\toprule
Model & RMSE (°C) & Gain (°C) & Frac.\ pos. \\
\midrule
Forcing only                                 & 0.787 & ---    & ---  \\
AR(1)                                        & 0.723 & +0.064 & 1.00 \\
AR(2) [best AR]                              & 0.695 & +0.093 & 0.96 \\
AR(3)                                        & 0.722 & +0.065 & 0.68 \\
Lag-weight filter, Atlantic only             & 0.782 & +0.005 & 0.75 \\
Two-driver finite lag                        & 0.659 & +0.128 & 0.86 \\
MA(5), Mediterranean [simple]               & 0.559 & +0.228 & 0.96 \\
EWMA $\phi{=}0.85$, Mediterranean [simple]  & 0.538 & +0.249 & 0.96 \\
Lag-weight filter, Med+Atl                  & 0.598 & +0.189 & 0.96 \\
Lag-weight filter, Mediterranean            & 0.578 & +0.209 & 0.96 \\
Med SST + lag-weight filter                 & 0.469 & +0.318 & 0.96 \\
\textbf{Med SST lag-0}                       & \textbf{0.449} & \textbf{+0.338} & \textbf{0.93} \\
\bottomrule
\multicolumn{4}{l}{\footnotesize
  Kernel: $\alpha=0.70$, $\theta=0.35$\,yr$^{-1}$,
  $\tau_{\rm mem}=2.9$\,yr (nested CV).
  MA(5) and EWMA: identical nested-CV protocol.}
\end{tabular}
\end{table}

The geographic pattern of filter gains is spatially coherent with the
persistence map.
The largest gains occur in the southeastern Mediterranean arc---TUR
($+0.43\,^{\circ}$C), MED-E ($+0.40\,^{\circ}$C), GRE
($+0.39\,^{\circ}$C)---where Mediterranean heat-storage memory is
most directly coupled to regional surface temperature.
The sole negative gain is GIC ($-0.09\,^{\circ}$C), the sub-region with
the highest Hurst exponent ($H=0.876$) and the deepest thermohaline
memory, where the relevant dynamics operate at timescales beyond the
1--5-year filter.
The per-region correlation between filter gain and $H$ is $r\approx-0.35$
($p\approx0.07$)---not statistically significant, but qualitatively
coherent: sub-regions with the very highest persistence are not well
served by a short-lag filter.

\subsection{Memory signal after circulation controls}
\label{sec:circ}

Adding NAO, AO and PNA to forcing-only reduces RMSE by only
$0.002\,^{\circ}$C ($0.787\to0.785\,^{\circ}$C).
The lag-weight filter applied after circulation control achieves
RMSE $=0.573\,^{\circ}$C, a gain of $+0.211\,^{\circ}$C, positive in
27/28 sub-regions (Table~\ref{tab:circ}).
Partial correlations after removing forcing and circulation:
$\bar{r}=0.407$ for the filter, $\bar{r}=0.573$ for Mediterranean
lag-0 SST, both positive in all 28 sub-regions.
The lag-memory contribution is not a proxy for the circulation modes tested.

\begin{table}[htbp]
\caption{Circulation-control model performance.
  Mean RMSE (°C) and partial correlations after removing forcing and
  NAO/AO/PNA effects (2006--2024 validation).}
\label{tab:circ}
\centering
\small
\begin{tabular}{lc}
\toprule
Model configuration & Mean RMSE (°C) \\
\midrule
Forcing only                             & 0.787 \\
Forcing + NAO/AO/PNA                     & 0.785 \\
Lag-weight filter, Mediterranean         & 0.598 \\
Lag-weight filter + NAO/AO/PNA           & 0.573 \\
Med SST lag-0                            & 0.449 \\
Med SST lag-0 + NAO/AO/PNA               & 0.455 \\
Med lag-0 + lag-weight + NAO/AO/PNA      & 0.470 \\
\midrule
\multicolumn{2}{l}{Partial $r$ (after forcing + circulation):} \\
\multicolumn{2}{l}{\quad Filter: $\bar{r}=0.407$;\quad
  Med lag-0: $\bar{r}=0.573$ (both positive in all 28 sub-regions)} \\
\bottomrule
\end{tabular}
\end{table}

\begin{figure}[htbp]
  \centering
  \includegraphics[width=\linewidth]{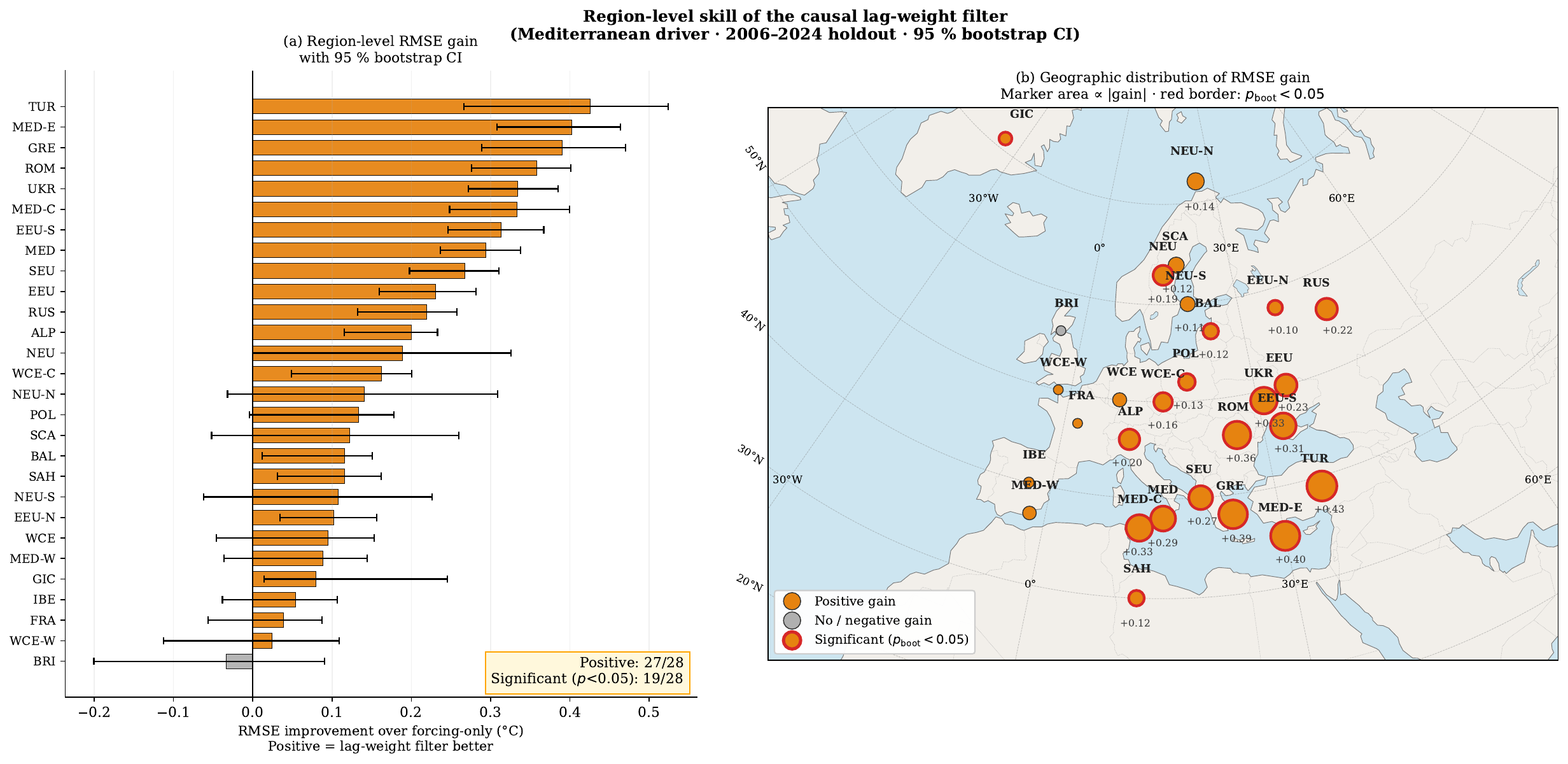}
  \caption{(a) Region-level RMSE gain of the lag-weight filter over
    forcing-only (2006--2024 holdout, 95\% bootstrap CI); positive in
    27/28 sub-regions, significant ($p_{\rm boot}<0.05$) in 21/28.
    (b) Geographic distribution; marker size $\propto|\text{gain}|$;
    red border: $p_{\rm boot}<0.05$.
    The pattern mirrors the Hurst map: largest gains in the Mediterranean
    arc; no gain in GIC (highest $H$, multi-decadal thermohaline memory).}
  \label{fig:regional_gain}
\end{figure}

\begin{figure}[htbp]
  \centering
  \includegraphics[width=\linewidth]{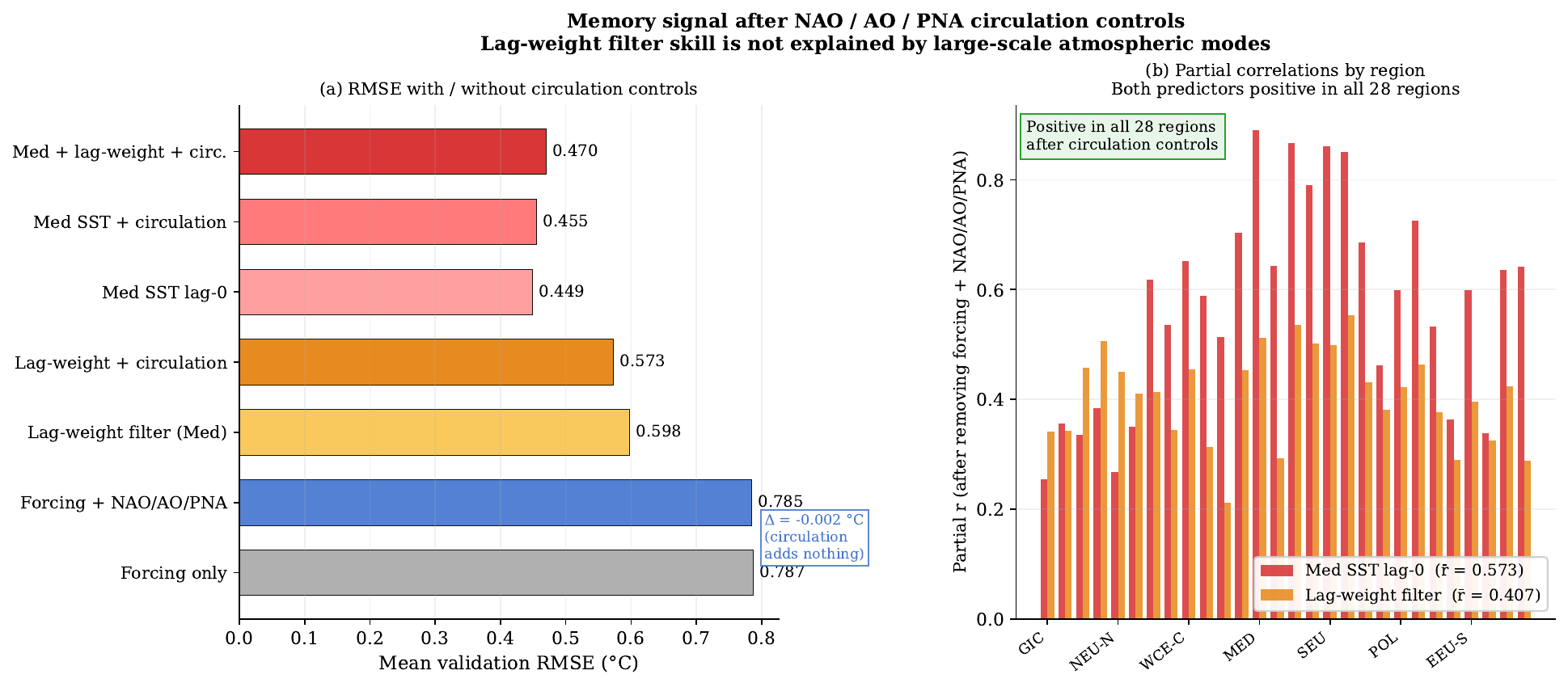}
  \caption{(a) RMSE with/without NAO/AO/PNA circulation controls;
    adding circulation to forcing gives negligible benefit
    ($\Delta=-0.002\,^{\circ}$C).
    (b) Partial correlations of the lag-weight filter (orange) and
    Med SST lag-0 (red) after removing forcing and circulation;
    both positive in all 28 sub-regions.}
  \label{fig:circulation}
\end{figure}

\section{Discussion}
\label{sec:discussion}

\subsection{The gap this study fills}

AMOC cold-blob dynamics have been documented in SST records
\citep{ceasar2018coldblob,rahmstorf2015exceptional} and linked to
European temperature extremes in model experiments \citep{buckley2016amoc}.
Mediterranean SST has been connected to European heat in case studies
and regional analyses \citep{giorgi2006hotspot,ionita2020med}.
Long-range dependence in individual climate series has been characterised
using DFA \citep{franzke2012,beran1994}.
The gap is the absence of an integrated, spatially resolved demonstration
that the interannual persistence pattern across 28 European sub-regions
constitutes a coherent spatial signature of AMOC thermohaline dynamics,
simultaneously organising warming-rate heterogeneity and short-lag
oceanic predictability.

This study fills that gap through four convergent lines of evidence:
(i) the DFA1 Hurst exponent maps onto the cold-blob-influenced structure;
(ii) the $H$--$\dot{T}$ gradient is robust to five null families and five
residualisation choices;
(iii) the signal localises in low-frequency interannual persistence,
not in trend curvature;
and (iv) the geography of lag-filter skill is qualitatively coherent
with the persistence map.
The progressive strengthening of the gradient in more recent warming
periods ($r=-0.550$ to $-0.875$ across 1981--2024 to 2000--2024) argues
against the result depending on a specific choice of reference period
and is consistent with intensifying AMOC weakening sharpening the
spatial contrast.

\subsection{Physical interpretation: oceanic buffering versus land-surface amplification}

The central empirical pattern should be read as a contrast between two physical
regimes rather than as a single-cause AMOC attribution. Atlantic-proximal
sub-regions combine high persistence with slower warming, consistent with oceanic
buffering by North Atlantic SST anomalies and the cold-blob region. This is the
part of the pattern most naturally connected to AMOC-related thermohaline memory.
Continental-interior sub-regions combine lower persistence with faster warming.
This is not a paradox. It is consistent with the well-established role of
soil-moisture--temperature feedbacks in European hot extremes
\citep{seneviratne2010soil,hirschi2011,miralles2014megaheatwaves}: when spring or
early-summer soil moisture is depleted, evaporative cooling weakens and the same
radiative background produces stronger near-surface warming. Such processes can
amplify warming rapidly while reducing the apparent interannual temperature
memory measured by annual DFA, because the realised anomaly depends strongly on
the current-year hydrological state.

Thus, the negative $H$--warming gradient is not interpreted as ``memory protects
against warming'' in a universal sense. Instead, it separates regions dominated
by slow oceanic thermal memory from regions dominated by faster land-surface and
circulation innovations. This distinction clarifies why high-$H$ regions warm
more slowly on average, while low-$H$ continental interiors can experience the
largest recent acceleration.

\subsection{How far the cold-blob interpretation can be taken}

The evidence presented here is strongest as a spatial consistency argument. The
high-$H$/slow-warming pole of the gradient is geographically aligned with
Atlantic-proximal sectors, the North Atlantic cold-blob literature, and direct
observations of recent AMOC weakening. The study also includes an Atlantic
cold-blob SST index in the predictor hierarchy. However, the analysis does not
estimate an ocean dynamical model and does not prove that AMOC weakening is the
unique cause of the land-temperature persistence gradient. Maritime heat capacity,
NAO-related variability, the East Atlantic pattern, blocking and regional SST
anomalies may all contribute. We therefore use the phrase ``consistent with North
Atlantic cold-blob influence'' deliberately. A stronger causal test would require
explicit covariance between land persistence and a dynamically constrained
cold-blob/AMOC index, ideally using longer SST records, ORAS5-type ocean
reanalyses or CMIP6 large ensembles. This is a priority for follow-up work.

\subsection{Honest assessment of the predictor hierarchy}

The dominance of the contemporaneous Mediterranean SST ($-43\%$ RMSE)
is consistent with prior regional analyses but is here placed in a
systematic 28-sector, strict-holdout framework for the first time.
The transparent comparison of the tempered kernel with MA(5) and EWMA
confirms that the operative memory is short interannual lag persistence
(lags 1--5\,yr): both simple alternatives achieve comparable or slightly
better validation RMSE.
This honest result strengthens rather than weakens the paper, because
it shows that the predictive signal is robust to the choice of filter
form and accessible with standard tools.
The tempered kernel's contribution is interpretive---its parameters
$\alpha$ and $\theta$ identify the 1--3\,yr window as the dominant
contributor, consistent with Mediterranean heat-storage timescales---but
the memory signal itself does not depend on the specific parameterisation.

The correlation between per-region filter gain and $H$ is not
statistically significant ($r\approx-0.35$, $p\approx0.07$) and is
presented as qualitative coherence rather than a quantitative claim.
The key qualitative observation---that the one sub-region with a
negative gain (GIC) is the one with the highest $H$ and the deepest
thermohaline memory---is internally consistent with the physical
interpretation but should not be interpreted as a formal result.

\subsection{Limitations and future directions}
\label{sec:limitations}

The analysis is restricted to annual mean temperatures, which cannot
resolve sub-seasonal circulation dynamics relevant for individual
heat waves or spring--summer soil-moisture feedbacks.
The circulation controls (NAO, AO, PNA) represent dominant extratropical
modes but do not include the East Atlantic pattern, the Scandinavian
blocking index, or JJA-specific circulation indices; including these
would strengthen the attribution by ruling out a broader range of
atmospheric mechanisms.
The spatial block permutation assumed four coarse geographic blocks;
a formal Moran's~I test or geostatistical permutation would provide
a more rigorous treatment of spatial autocorrelation.
The RAPID record (21\,yr) is too short for direct multi-decadal AMOC
calibration; extending the analysis to the Caesar--Boers AMOC proxy
(1870--present), ORAS5 ocean reanalyses or CMIP6 large ensembles would allow
testing the persistence gradient over a longer window and provide more direct
AMOC validation than the cold-blob SST proxy used here. The current manuscript
therefore establishes a robust spatial signature consistent with the cold-blob
mechanism, not a closed causal attribution.
The $H$--$\dot{T}$ association is spatial and correlational; causal
attribution to AMOC weakening, rather than to other correlated slow
processes, would require targeted modelling or longer-timescale
observational validation.

\section{Conclusions}
\label{sec:conclusions}

This study documents two connected empirical contributions for 28
IPCC\,AR6 European sub-regions using ERA5 data (1950--2024), with the
principal finding in the spatial persistence structure and the
supporting finding in the annual predictor hierarchy.

The principal contribution is a robust negative spatial association
between the DFA1 Hurst exponent and the 1996--2024 warming rate
($r=-0.792$, $p=5.1\times10^{-7}$, $n=28$).
Atlantic-proximal sub-regions with high interannual persistence warm
slowly; continental-interior sub-regions with low persistence warm
rapidly.
This gradient survives five residualisation strategies, three
independent memory estimators, leave-one-region-out analysis, and five
independent null-test families (all $p\leq0.001$), including a spatial
block permutation.
It strengthens progressively in more recent warming periods
($r=-0.875$ for 2000--2024), consistent with intensifying AMOC
thermohaline weakening.
It disappears under DFA2, localising it in the low-frequency
interannual-to-decadal persistence generated by thermohaline memory,
not in short-scale variability or trend curvature.
Together, this multi-line evidence constitutes an observational spatial
signature consistent with North Atlantic cold-blob influence and thermohaline
memory contributing to the heterogeneity of European warming acceleration, while
leaving room for continental land--atmosphere feedbacks and circulation variability
as complementary mechanisms.

The supporting contribution is a systematic annual predictor hierarchy
under a strict 19-year holdout.
The contemporaneous Mediterranean SST dominates, reducing RMSE by 43\%.
A causal lag-weight filter provides a secondary independent contribution,
outperforming AR(2) by $0.12\,^{\circ}$C and surviving circulation
controls.
Comparison with simple moving averages and exponentially weighted
averages---evaluated under the identical protocol---shows that the
operative memory is short interannual lag persistence at lags 1--5\,yr,
accessible with simple tools.
The geographic pattern of filter gains mirrors the persistence gradient,
confirming internal consistency of the physical interpretation.

Sub-regions buffered by strong thermohaline memory warm slowly on
average, but their persistence makes them susceptible to multi-year
clusters of anomalously warm years that mean-trend projections
systematically underestimate.
Persistence-aware risk characterisation---based on the $H$ spatial
gradient documented here---is recommended for risk assessments in
Atlantic-proximal European sectors.

\section*{Data and code availability}

ERA5: \url{https://cds.climate.copernicus.eu} \citep{hersbach2020era5}.
RAPID: \url{https://rapid.ac.uk/data/data-download} \citep{moat2020rapid}.
NOAA/CPC: \url{https://www.cpc.ncep.noaa.gov}.
IPCC\,AR6 regions:
\url{https://github.com/SantanderMetGroup/IPCC-Atlas}
\citep{iturbide2020regions}.

Code and tables:
It will be archived soon in the GitHub repository
\url{https://github.com/mauricio-herrera/europe-memory-warming}
(DOI: 10.5281/zenodo.XXXXXXX).

\section*{Author contributions}
M.H.-M.\ conceived and designed the study, performed all analyses
and wrote the manuscript.
A.G.-F.\ and D.R.\ contributed to interpretation and revision.

\section*{Competing interests}
The authors declare no competing interests.

\section*{Acknowledgements}
ERA5, RAPID--MOCHA--WBTS and NOAA/CPC teams for publicly available data.

ANID/ANILLO ATE250004.

\bibliography{references}
\end{document}